\documentclass[aps,prd,preprint,superscriptaddress,preprintnumbers,nofootinbib]{revtex4-1}

\usepackage{amsmath}
\usepackage{amsfonts} 
\usepackage{amssymb}
\usepackage{graphicx}
\usepackage{hyperref}
\usepackage{tensor}
\usepackage{siunitx}
\usepackage{physics}

\usepackage{appendix}
\newcommand{\be}{\begin{equation}}
\newcommand{\ee}{\end{equation}}

\newcommand{\beq}{\begin{equation}}  \newcommand{\eeq}{\end{equation}}
\newcommand{\beqn}{\begin{eqnarray}}
 \newcommand{\eeqn}{\end{eqnarray}}
\newcommand{\bal}{\begin{aligned}}   \newcommand{\eal}{\end{aligned}}
\newcommand{\bea}{\begin{eqnarray}}  \newcommand{\eea}{\end{eqnarray}}
\def\d{\mathrm{d}}

\hypersetup{colorlinks=true,
			urlcolor=purple,
			citecolor=blue,
			linkcolor=magenta,}

\begin{document}

\title{\Large Toroidal $\&$ orbifold compactifications at large D \newline
and  D-duality}
\preprint{LMU--ASC 25/21}
\preprint{MPP-2021-112}

{~}

\author{\large Dieter~L\"ust}
 \affiliation{Arnold-Sommerfeld-Center for Theoretical Physics, Ludwig-Maximilians-Universit\"at, 80333 M\"unchen, Germany\\[1.5ex]}
\affiliation{Max-Planck-Institut f\"ur Physik (Werner-Heisenberg-Institut),
             F\"ohringer Ring 6,
             80805, M\"unchen, Germany}

\vspace{0.4cm}

\begin{abstract}
\vspace{0.3cm}
\noindent
In this paper I will further investigate the spectrum of quantum gravity or string theories at large  number  of dimensions. We will 
see that volumes of certain orbifolds shrink at large D. It follows 
that
the mass spectra of the leading Kaluza-Klein towers and also of wrapped brane states on these orbifolds
possess a non-trivial dependence on D: KK modes become heavy at large D, whereas wrapped branes become light.
This observation can be used either to apply the Large Dimension Conjecture  or, as we will do,  to investigate the possibility of a D-duality symmetry, which relates in gravity compactifications  of different dimensions.
We will  set up the general rules for D-duality in higher dimensional gravity. However due to existence of the critical dimensions, D-duality is quite restricted in string theory.  As simple tests
for D-duality in string theory, we will discuss the duality between M-theory on a two-dimensional orbifold, namely the 
M\"obius strip, and an IIB S-fold on a circle, which corresponds to the heterotic CHL string, as well the duality between a (truncated) 12-dimensional theory on a three-dimensional
orbifold and another IIB S-fold  compactification. Finally I also comment on the possible existence of exotic theories at large D.

 \vspace{1cm}
\end{abstract}

\maketitle

\tableofcontents
%\break

\section{Introduction}

String theory provides an explicit (and possibly unique) realization for a theory of quantum gravity and 
the existence of a critical dimension $D=10(26)$ is an essential and characteristic feature of the (bosonic) super string theory \cite{Goddard:1973qh,Schwarz:1974ix,Goddard:1972iy,Polyakov:1981rd}.\footnote{Some non-critical
string constructions were considered in \cite{Myers:1987fv,Antoniadis:1988aa,Maloney:2002rr,Aharony:2006ra,McGreevy:2006hk}.}
The requirement for a critical dimensions originates from the consistency of the underlying world-sheet conformal field theory as well as from Lorentz invariance of the target space theory. 
In addition it is known \cite{Nahm:1977tg,Cremmer:1978km}
for many years that local supersymmetry in flat space-time can be only realized in supergravity theory with maximal number of dimensions $D=11$. 
Nevertheless, besides these arguments from string theory or from supersymmetry, it would be very interesting to get bounds on the number of space-time dimensions from
general considerations in quantum gravity.

On the other hand, the concept of dimension of space-time appears not completely well-defined in string theory or in quantum gravity.
In particular it is known that gravitational theories in different numbers of space-time dimensions can be related by duality symmetries.
E.g. the type II A superstring in 10 space-time dimensions theory was shown to be equivalent to 11-dimensional M-theory 
\cite{Witten:1995ex}
and 10-dimensional IIB superstring theory
can be formulated in terms of F-theory \cite{Vafa:1996xn}, 
which possesses several  features of a 12-dimensional theory. In addition via the AdS/CFT correspondence \cite{Maldacena:1997re}
it was realized that a quantum gravity theory
in $d+1$ dimensions can be equivalently described by a d-dimensional quantum field theory without gravitational degrees of freedom.

Another interesting aspect, relevant for the question about the number of space-time dimensions, is the possibility of emergent space-time and emergent geometry.
This question can be nicely addressed within the socalled swampland scenario
     \cite{Vafa:2005ui,Ooguri:2006in,Palti:2019pca,vanBeest:2021lhn}.
In quantum gravity or in 
string theory, objects revealing the geometric nature of the theory, or its 
stringy nature, can become part of the low-energy modes, as happens for instance in certain infinite distance limits in field space, like  the  decompactification limit  \cite{Heidenreich:2018kpg,Grimm:2018ohb}   or the  tensionless 
string limit  \cite{Lee:2019wij}. 
It is for example well-known that the presence of towers of Kaluza Klein modes in an effective field theory is a clear signal for the emergence of  an additional  compact geometry.
It has been also argued in the context  of the AdS distance conjecture \cite{Lust:2019zwm}
 that quantum gravity in pure AdS space-time belongs to the swampland, which means that in the limit of large AdS radius always a tower 
of light states is emerges, which corresponds to the existence of a additional space-time dimensions.

In a recent paper \cite{Bonnefoy:2020uef}, also motivated by ideas from the large D expansion of gravity \cite{Emparan:2013moa,Emparan:2020inr}, we  addressed the question
is large $D$ quantum gravity in the swampland?
Namely we
have provided certain swampland arguments to constrain the 
number of space-time dimensions coming from the spectrum of Kaluza-Klein modes, from black hole physics and from the weak gravity conjecture.
Concretely, we proposed in \cite{Bonnefoy:2020uef} to include the number $D$  of space dimensions in the parameter space
of quantum gravity. This amounts to treat $D$, in addition to the geometric parameters like the radius $R$ of a compact space, as new swampland parameter and to include the dependence on $D$
into the distance functionals $\Delta(D)$, which measure the distances between different backgrounds in quantum gravity. 
Then, 
the proposed  {\sl Large-D Conjecture (LDC)},  together with a negative distance conjecture (NDC)
state that $\Delta(D)$ for certain towers of states should stay positive as a function of the number of space-time dimensions, unless there is a dual tower of states at large space-time dimensions.
For the Kaluza-Klein states this is equivalent to require that the leading Kaluza-Klein tower is always lighter than the effective Planck scale of the theory.
As a general consequence of the LDC,
the number of space-time dimensions has to be smaller or equal to some critical value that depends on the typical size of space-time.
Furthermore in \cite{DeBiasio:2021yoe},
we considered  new geometric flow equations, called D-flow, which describe the
variation of space-time geometries under the change of the number of dimensions.

In \cite{Bonnefoy:2020uef}
we also briefly mentioned a possible new D-duality symmetry between large and small dimension.\footnote{D-duality, although in a different setting, was also discussed in \cite{Green:2007tr}.
Moreover c-duality was discussed in \cite{Hellerman:2007fc}.}
D-duality states that for certain towers that are light in $D$ dimensions and heavy   in backgrounds of dual number dimensions, denoted by $\tilde D$,
there should exist dual towers, which are heavy  in D dimensions and light  in  $\tilde D$ dimensions, and vice versa.
For example the  light  tower in D-dimensions and the heavy tower in $\tilde D$ dimensions can be given by perturbative KK modes, whereas 
their dual towers
are given in terms  of non-perturbative dual winding modes.
For D-duality to be a symmetry of gravity, the spectra (and interactions) in $D$ and in $\tilde D$ dimensions should agree with each other.

It is the aim of this paper to investigate in more detail the spectrum of gravity theories at large D, where in this paper, D will always refer to the number of internal compact dimensions.
In \cite{Bonnefoy:2020uef} we have shown  the Kaluza-Klein spectrum of $AdS_d\times S^{D}$ has a non-trivial dependence on the number of dimensions,
since the volume of the D-dimensional sphere depends on the number of space-time in a particular way. Namely keeping the diameter $R$ of the sphere, which defines the maximal distance between two points fixed,
the volume of the sphere nevertheless shrinks 
in the limit of large D. This means that the KK modes become heavy for fixed $R$ and large D.
We have utililzed this observation for the LDC in order to constrain the number of dimensions for $AdS_d\times S^{D}$ geometries by $R$ of the D-dimensional sphere. 
However  for  the curved $AdS_d\times S^{D}$ backgrounds
 the KK masses in AdS space start to become heavier than the Planck mass in the regime where the space-time curvature is already large.

Here we  continue this discussion by  analzying the Kalazu-Klein spectra for flat backgrounds, where there are no curvature corrections, namely for tori $T^{(D)}$ and for certain orbifolds ${\cal O}^{(D)}$
at arbitrary dimensions $D$.
We will see that, whereas the leading KK tower of tori $T^{(D)}$ is independent of $D$, the leading tower of certain D-dimensional orbifolds ${\cal O}^{(D)}$
indeed possesses an interesting and non-trivial dependence on D:
like for the higher-dimensional spheres, keeping the diameter $R$ of the orbifold fixed, 
the orbifold volume is also sensitive to the number of space-time dimensions and shrinks at large D. So the KK modes of the orbifold in question become heavy at large D, when keeping the diameter $R$ 
of ${\cal O}^{(D)}$ fixed.
This opens the possibility to apply the LDC for orbifold compactifications and raised the question if one can derive bounds on D for toroidal and orbifold compactifications.
In fact we will see that, unlike for the sphere $S^{D}$, the topology of the orbifolds ${\cal O}^{(D)}$ is such that it also allows, at least in principle, for a full tower of dual states, which is due branes that are completely wrapped around the orbifold.
Then the  mass scale of the wrapped brane states on the orbifolds possess a non-trivial dependence on D,  namely they become light in the large D limit, opposite to the mass scale of the KK states. With this observation 
we will investigate D-duality symmetries for  certain
orbifold backgrounds in gravity and in string theory.
%We will do this by analyzing the Kalazu-Klein spectra for tori and certain orbifolds at arbitrary dimensions, focussing in particular on those KK modes, with masses that exhibit a non-trivial dependence of $D$, which 
%are light for small $D$ and which become heavy for  larger $D$. Dual to these states there will be certain wrapped brane modes with precisely opposite behaviour.
As we will discuss, an already known example of this kind of D-duality is the duality between type II superstring on $S^1$ and M-theory on $T^2$, or some orbifolds version of it.
Also  certain F-theory "compactifications" fall in the class of D-dual theories.

The paper is organized as follows. 
In the next section, we will briefly review the NDC and the LDC and the motivation for D-duality.
Then in section 3, we will extend the discussion of     \cite{Bonnefoy:2020uef}     by analyzing the leading KK spectra for toroidal and two classes of orbifold backgrounds at arbitrary $D$.
 %as well as the relevant dual winding spectrum.
We will also discuss the topological properties of the D-dimensional orbifolds. 
These will allow the existence of a tower of wrapped brane states, which can  at least in principle be dual to the tower of KK states. As necessary conditions for D-duality, 
We will set up the D-duality transformation rules, namely  as relations between  $D$ and $\tilde D$ as a function of the radius $R$ of the orbifold.
We will also discuss the general rules for D-dual pairs of  orbifold spaces. 
 However, as we will discuss,  these examples can realized in string theory
 only in a quite restricted setting, namely as string theory/M-theory dualities, or as string theory/F-theory dualities. In the latter case, 
 a certain truncation of the spectrum is
necessary.

\section{The LDC and D-duality}

In this section we recall briefly some of the main aspects of the Large Dimension Conjecture, which was introduced  in 
\cite{Bonnefoy:2020uef}.
The starting point is the {\sl Swampland Distance Conjecture} (SDC) \cite{Ooguri:2006in}, which
states that  at large distances $\Delta$ in the field space 
of a $d$-dimensional (effective) quantum gravity theory there must be an infinite tower of states with mass scale $m$ such that 
\be
SDC:\quad m = M_P e^{-\Delta } \,.
\label{dsc}
\ee 
The mass scale $m$ can be seen as the scale below which the EFT provides a good description of 
the low-energy physics. Hence the d-dimensional EFT breaks down when $\Delta\rightarrow\infty$. 

In general $\Delta$ determines the geodesic distance in the parameter space of background in quantum gravity, which are characterized by some parameters $\phi$, 
implying that $\Delta$ is a certain function of $\phi$, i.e. $\Delta=\Delta(\phi)$. In the EFT $\phi$ is in general associated to a canonically normalized scalar field $\Phi(\phi)$ with kinetic energy
${\cal L}_{EFT}\simeq{1\over 2}(\partial\Phi)^2+\dots$.
Then the distance functional is proprtional to $\Phi$:
\begin{equation}
\Delta(\phi)=\lambda\Phi(\phi)\,.\label{falloff}
\end{equation}

For (string) compactifications the SDC is due to the higher dimensional nature of the theory, namely the relevant tower of states is given in terms of KK momentum states with masses $m_{KK}=1/R$. Then has the form
$\Delta_{KK}\simeq\log R$ and becomes large in the decompactification limit $R\rightarrow\infty$. Another realization of the SDC is due to the tower of string exciations that become massless in the weak
coupling limit, i.e. $\Delta_{string}\simeq-\log g_s$.

On the other hand in the opposite limit $R\rightarrow0$ the KK modes are heavy and the geometrical picture of a compact space gets lost.
However in string theory there is often a duality symmetry, like T-duality with a 
T-dual tower of fundamental string winding modes (F1-strings) with masses $m_{wind}=R$ and distance functional $\Delta_{wind}=-\log R$, which becomes large for $R\rightarrow0$.
More generally, duality symmetries often follow from the SDC, when considering "opposite" large distance limits in field space.

The SDC implicitly also implies that if it happens that there is an infinite distance limit in field space without a corresponding light tower of states that satisfies the SDC, then it cannot be possible to approach this infinite distance point.
Based on this statement, we now reverse the logic 
and define, for a given tower tower of states $\left|i\right>$, the quantity $\Delta_i$ (the ``distance'') as the negative logarithm of its typical mass scale $m_i$, i.e.
\begin{equation}
\Delta_i \sim - \log  m_i\, .
\end{equation}
In addition there possibly can be  a dual tower of states $\left|\tilde \imath\right>$ with the dual distance $\widetilde \Delta_i \sim - \log \widetilde m_i$.

On the basis of these definitions and arguments we now formulate the  {\sl Negative Distance Conjecture} (NDC) demanding
that,  for a particular leading tower $\left|i\right>$,  an "infinite distance" limit $\Delta_i\rightarrow -\infty$ is obstructed, i.e.~not allowed,
unless there is a dual tower $\left|\tilde \imath\right>$, which becomes light in this limit. 

\vskip0.3cm
\noindent
There are three basic arguments in favour of the NDC: 

\vskip0.3cm

\noindent (i) For a tower with negative $\Delta_i$, the associated tower mass scale is above the Planck mass, $m>M_P$. For the leading KK tower this obstructs a proper geometric interpretation of the tower.

\noindent (ii) For a tower with negative $\Delta_i$,  the coupling constant $g_i$, like $g_{KK}$ or $g_s$, associated to the tower $\left|i\right>$ becomes large, i.e. $g_i>1$. So the EFT becomes strongly coupled and hard
to be controlled.

\noindent (iii) For a tower with negative $\Delta_i$,  the effective Planck mass $M_P^{(d)}$  in d dimensions becomes smaller than the fundamental Planck mass $M_P^{(D)}$ or becomes smaller than the string scale
$M_s$, which means the effective gravity theory becomes strongly coupled at energies lower than $M_P^{(d)}$.

\vskip0.3cm
So these are three circumstances, which should not happen in a well defined and controllable EFT for quantum gravity, unless there is a weakly coupled dual tower of states, with  $\tilde m<M_P$ and $\tilde g_i<1$.

Typically the tower masses depend on the geometric parameters $\phi$, i.e. $m=m(\phi)$.
A good example for the NDC is given by the 
 KK modes, where the limit $\Delta_{KK} \rightarrow -\infty$ implies $\phi \rightarrow -\infty$ and the SDC resp. NDC says that there must be a second tower of states or $\phi \rightarrow -\infty$ cannot be possible. A further example of the NDC in string theory is provided by the string states (i.e. by the string itself) and the associated tower of states in the limit of weak string coupling, $g_s\to 0$. Indeed, the string scale reads
\begin{equation}
M_s^2=(g_s)^{1/2}M_p^2\, \xrightarrow[g_s\to 0]{} 0 \ .
\end{equation}
The NDC is then a statement about the the strong coupling limit $g_s\to\infty$ limit. Naively, this strong coupling limit is not controllable, however string dualities allow to probe this limit, where a new tower of light states arises, provided e.g.~by (wrapped) NS5-branes or by D1-strings in the type IIB string. Actually, it has been argued \cite{Lee:2019wij} that those two cases of KK modes and string states may be everything there is at infinite distance limits, as long as a suitable duality frame is chosen to analyse the limit.

In addition the tower masses  may also depend on the number D of compact
dimensions in a non-trivial way: $m=m(\phi,D)$. Therefore we want to include D into the swampland distance functional as new parameter:
\begin{equation}
\Delta_i=\Delta_i(\phi,D)\, .
\end{equation}
This is unusual, since there is no known dynamical theory for D, and also no kinetic term for D in the EFT. 
But the general idea is that $\Delta_i(\phi,D)$  determines the distance between space-time geometries of different dimension.

\vskip0.3cm
Having introduced D as swampland parameter, the {\sl Large Distance Conjecture} (LDC) can now be formulated as an extension of the NDC:
If there is a  leading tower of states $\left|i\right>$ with a non-trivial D-dependence,  the corresponding distance must be a positive function of the EFT fields $\phi$ and of the number of dimensions D,
\begin{equation}
\Delta_i(\phi,D)\geq 0\, ,
\end{equation}
unless there is a dual u tower $\left|\tilde \imath\right>$, such that $\tilde \Delta_i(\phi,D)$ becomes positive when $\Delta_i(\phi,D)$, when $\Delta_i(\phi,D)$ changes its sign.
Since $\Delta_i(\phi,D)$ depends on $\phi$ and on D, the LDC normally puts a bound on D as a function of $\phi$, like:
\begin{equation}
D\leq D_0(\phi)\, .
\end{equation}

Let is briefly recall that for a compactification of the form $M_{d} \times K_{D}$ with arbitrary numbers $d$ and $D$ of non-compact and compact dimensions ($D_{tot}=d+D$) and metric of the form,
\begin{equation}
\d s^2_{D_{tot}} = e^{2 \beta \varphi} d s^2_{M_{d}} + e^{2 \phi} d s^2_{K_{D}} \,,
\label{ansatzdTimesd'}
\end{equation}
($\beta=-{D\over d-2}$) the KK mass scale is given as \cite{Bonnefoy:2020uef}
\beq\begin{aligned}
m_{KK}\approx {\cal V}^{-{1\over D}}_{D}M_P^{(d+D)} \approx M^{(d)}_P{\cal V}_{D}^{-{d+D-2\over D(d-2)}} 
= M^{(d)}_P \exp\left\lbrack-\sqrt{\frac{d+D-2}{D(d-2)}}\Phi\right\rbrack\, .
\label{KKmasses}
\end{aligned}\eeq
Here $M_P^{(d+D)}$ and $M^{(d)}_P$ are the Planck masses in the (d+D)-dimensional or d-dimensional Einstein frames, respectively, and ${\cal V}_{D}$ is the volume of the compact space measured in units
of $1/M_P^{(d+D)}$.
It follows that the constant $\lambda(d,D)$ in eq.(\ref{falloff}) is given by
\begin{equation}
\lambda(d,D)=\sqrt{\frac{d+D-2}{D(d-2)}} \,.
\end{equation}
For $d=D=D_{tot}/2$, this fall-off parameters goes to zero in the large D limit: $\lambda\approx\sqrt{1\over D}\rightarrow 0$.
In addition also the canonical field $\Phi$ may depend on the number of dimensions and the LDC bound  for KK modes reads
\begin{equation}
\Phi(d,D)\geq 0\, ,
\end{equation}

Let us now give some qualitative arguments 
under which circumstances  we expect the LDC bound to apply and where, on the other hand,  dual tower of states are in principle possible.
Actually there are two kinds of generic situations: 

- there are no fluxes and no potential for the volume modulus: these backgrounds are typically tori, orbifolds or Calabi-Yau spaces without fluxes. As we will discuss, for certain
orbifolds the KK masses have a non-trivial D-dependence. Furthermore 
dual towers of states can be present and can lead to a D-duality-symmetry.  

- there are non-zero fluxes: these backgrounds are typically $AdS_d \times S^{D}$ or warped Calabi-Yau spaces with fluxes. Here the complex structure moduli are typically frozen out  and flux quantization does not allow for small K\"ahler moduli.
Dual towers are not present because of the topology of the sphere.
   As it was discussed \cite{Bonnefoy:2020uef},
   in this case the LDC  provides an upper bound on D. The microscopic explanation of the bound on D is related to flux compactification of the underlying D-brane model.

\vskip0.3cm
In the following we will further discuss the KK spectrum of tori and certain orbifolds at large D and  how D-duality symmetries can be possibly realized.

%%%%%%%%%%%%%%%%%%%%%%%%%%%%%%%%%%%%%%%%%%%%%%%%%%%%%%%%%%%%%%%%%%

\section{Toroidal and orbifold compactifications at arbitrary and large D}

In this section we analyze the spectrum of tori and certain orbifolds at arbitrary D, where this discussion 
 is a priori not restricted to string theory.

%\subsection{Distances in a fixed number of dimensions}

\subsection{Circle compactification}

Let us  consider the well known case of Kaluza-Klein modes on tori. First we will consider 
 the Kaluza-Klein tower of states in flat $d$-dimensional Minkowski space, which arises from compactification of an additional
internal circle of radius $R$.  So the total number of dimensions is $D_{tot}=d+1$.
In the string frame or respectively in the $(d+1)$-dimensional Einstein frame
their masses  are given 
\be
m_{KK}^2 = \biggl({l\over R}\biggr)^2=\biggl({l\over r}\biggr)^2 (M_P^{(d+1)})^2\, , 
%{\color{red}m_{KK}^2 = \biggl({m\over R}\biggr)^2}
\label{KK}
\ee 
 where the dimensionless radius $r=R/M_P^{(d+1)}$ is measured here and in the following in units of the $D_{d+1}$-dimensional Planck mass $M_P^{({d+1})}$.
The integer $l$ denotes the KK charge.

When the radius $r$ approaches  the critical radius $r_0=1$, the
KK tower becomes heavy and for $r<1$ the tower starts above the Planck scale. 
However in string theory, for $r<1$
the elementary string, with string tension 
\begin{equation}
T^{(1)}=M_s^2=(g_s^{(10)})^{1/2}(M_P^{(10)})^2\, ,
\end{equation}
where $g_s^{(10)}$ is the 10-dimensional string coupling constant and $M_s$ is the string mass, 
can wrap around the circle and the winding modes 
become  light. (Here we assume that $D_{tot}=10$.)
Specifically their masses are given as
\be
m_{wind}^2 
%= (M_P^{(D)})^2{m^2 R^2}
=n^2R^2M_s^4=n^2{r}^{2}g_s^{(10)}(M_P^{(10)})^2\, ,
\label{winding}
\ee 
where $n$ is the winding number of the string.

As it is well know, the T-duality symmetry is exchanging the KK with the winding spectrum, keeping the KK and winding masses invariant and acting in the following way on the other quantities:
\begin{equation}
l\leftrightarrow n\, ,\quad R\leftrightarrow 1/{(RM_s^2)}\, .\
\end{equation}
%{\bf Precise T-duality rules on R, Planck masses and string coupling and on r}
Here $R_0=1/M_s$ is the fixed point of the T-duality symmetry. 
T-duality also acts in a non-trivial way on the 10-dimensional string coupling constant,
\begin{equation}
g_s^{(10)}\quad\longrightarrow\quad {g_s^{(10)}\over R M_s}\,,
\end{equation}
and it follows that the Planck mass transforms under T-duality in the following way:
\begin{equation}
M_P^{(10)}\quad\longrightarrow\quad M_P^{(10)}(RM_s)^{1/4}\,.
\end{equation}

For $R>R_0$ the KK modes are building-up space-time geometry, whereas for $R<R_0$ the winding modes take over and are building-up a dual geometry.
Note that in the bosonic or heterotic string theories, T-duality symmetry acts as self-duality between KK modes and winding modes,
and the critical radius $R_0$ is the limiting radius, in the sense that one can restrict $r > R_0$ in the physical moduli space. Here the geometries at large and at small radii are completely equivalent to each other.
On the other hand, T-duality is mapping the KK spectrum of type IIA on the winding spectrum of type IIB and vice versa. Moreover the KK and winding spectra in type IIA or, respectively in type IIB, are not equivalent to each
other \cite{AbouZeid:1999fv}.
Therefore the moduli spaces for type IIA and for type IIB are given by all real values for the radii.
Staying e.g. within type IIA (or within type IIB), the KK geometry at large radii and the dual winding geometry are not completely equivalent to each other; however the KK geometry geometry of type IIA 
and the winding geometry of type IIB are indeed equivalent.
We will come back to similar issues in section \ref{Dduality}, when we discuss some other kind of KK and winding dualities that depend also on the number of space-time dimensions.

At the end of this section, we want to express the above quantities also in terms of the lower-dimensional Planck mass $M_P^{({9})}$ of the effective field theory in $d=9$ dimensions.
As it is well  known, $M_P^{({9})}$ and  $M_P^{({10})}$ are related as (all these relations can be generalized to arbitrary dimensions):
\begin{equation}
(M_P^{({9})})^7=R(M_P^{({10})})^8=r(M_P^{({10})})^7=r(g_s^{(10)})^{-7/4}M_s^7\,. 
\end{equation}
It follows that the KK masses and the winding masses in terms of the lower-dimensional Planck mass $(M_P^{({9})})$ can be expresses as
\begin{eqnarray}
m_{KK}^2& =&l^2 r^{-16/7} (M_P^{(9)})^2\, ,\nonumber\\
m_{wind}^2 
%= (M_P^{(D)})^2{m^2 R^2}
&=&n^2{r}^{16/7}g_s^{(9)}(M_P^{(9)})^2\, .
\end{eqnarray}
Here the 9-dimensional string coupling constant is defined as
\begin{equation}
(g_s^{(9)})^2={(g_s^{(10)})^2\over RM_P^{(9)}}\,.
\end{equation}
Note that the 9-dimensional quantities $M_P^{(9)}$ and $g_s^{(9)}$ are invariant under T-duality transformations.
Hence
T-duality is also manifest in the lower dimensional Einstein frame.

%U(1) GAUGRE GROUPS

\subsection{Two-torus}

Now we consider  the different Kaluza-Klein towers of states, which arises from compactification 
on a two-dimensional torus $T^{(2)}$, i.e. 
the total number of dimensions is $D_{tot}=d+2$.
A string compactification of a two-torus with metric $G_{ij}$ and
antisymmetric tensor field $B$ is described by four moduli fields, which can be grouped into two complex parameters, namely into a K\"ahler parameter $T$ and a complex structure parameter $U$
in the following way:
\begin{eqnarray}
T&=&(\sqrt G+iB)\, ,\nonumber\\
U&=&{1\over G_{11}}(\sqrt G-iG_{12})={r_2\over r_1}e^{-i\theta}\, .
\end{eqnarray}
The metric can be explicitly given in term of two radii $R_1$ and $R_2$ and one additional angle $\theta$ as
\begin{equation}
G_{11}=r_1^2\, ,\,\,\,G_{22}=r_2^2\, ,\,\,\,G_{12}=r_1r_2\sin\theta
\end{equation}
 and 
\begin{equation}
{\cal V}(T^{(2)})=\sqrt G=r_1r_2\cos\theta
\end{equation} 
denotes the overall volume of $T^{(2)}$.

The corresponding KK and winding spectrum is characterized by two KK numbers $l_i$ and two winding numbers $n_i$ and the corresponding masses are given as
\begin{equation}
m_{n_i,m_i}^2={|l_2-il_1U+in_1T-n_2TU|^2\over \Re T\Re U}\, .
\end{equation}

As it is well known, the KK and winding spectrum is invariant under discrete $O(2,2,{\mathbb Z})$ duality transformations.
These transformations can be written as $SL(2,\mathbb Z)_T\times SL(2,\mathbb Z)_U$, which act separately on the two moduli $T$ and $U$ as 
\begin{equation}
T\rightarrow {aT-ib\over icT+d}\, ,\quad U\rightarrow {a'U-ib'\over ic'U+d'}\, .
\end{equation}
Here the $SL(2,\mathbb Z)_T$ transformations generate the shifts of $B$ and the T-duality transformations and the $SL(2,\mathbb Z)_U$  transformations 
correspond to the geometric transformations that leave the 2-torus invariant.
In order to keep the spectrum unchanged, one has to transform the KK numbers $l_i$ and  winding numbers $n_i$ in an appropriate way.

Let us first consider the different KK towers, i.e. setting the winding numbers $n_1=n_2=0$.
In this case the spectrum is invariant under the geometric $SL(2,\mathbb Z)_U$ transformations.
In 
 the $(d+2)$-dimensional Einstein frame
the KK masses  take the following values
\be
m_{KK,l_1,l_2}^2 ={|l_2-l_1{r_2\over r_1}(\sin\theta-i\cos\theta)|^2\over r_2^2\cos^2\theta}(M_P^{(d+2)})^2\, .
\label{KKT2}
\ee 
%Note that all KK masses are positive definite.

For a rectangular  torus  with 
 $\theta =0$ and  the two radii set to be equal to each other, $r_1=r_2=r$,
the two leading KK towers $|l_i\rangle $ on the torus are the ones  with a single non-vanishing KK charge $l_i$ ($i=1,2$); they   possess masses
\be
|l_i\rangle :\quad m_{KK,l_i}^2 \equiv l_i^2m_{KK,(i)}^2(M_P^{({d+2})})^2\,,\quad m_{KK,(i)}^2={1\over r^2}={1\over {\cal V}({ T}^2) }\, .
\label{KKT2a}
\ee 
These two leading towers KK towers have a mass$^2$ scale identical to the inverse volume of the torus, and they
transform in the fundamental $\underline 2$ representation of $SL(2)$.
In order for the leading mass scale $m_{KK,(i)}^2 \leq 1$, one has to require that
\begin{equation}
r^2\geq {1}\, .
\end{equation}
Note that for $\theta\neq 0$, $m_{KK,(i)}^2$ is not anymore identical to the inverse volume of the torus.

\vskip0.3cm
Next, consider  the  subleading tower of KK states  $|l\rangle $ with $l_1=l_2=l$. Their masses  for $\theta=0$ are
\be
|l\rangle :\quad m_{KK,l}^2         =   l^2 m_{KK,(S)}^2(M_P^{({d+2)}})^2\,,\quad m_{KK,(S)}^2={2\over r^2}={2\over {\cal V}({ T}^2) }\, .
\label{KKT2dl}
\ee
This subleading KK tower transforms as a singlet under $SL(2)$.
In order for  the subleading, singlet mass scale $ m_{KK,(S)}^2\leq 1$, one has to require that
\begin{equation}
r^2\geq {2}\, .
\end{equation}

For the elementary string winding states the situation is analogous. On the rectangular torus with $r_1=r_2=r$, the $SL(2)$ doublet winding states with a single winding number $n_i\neq 0$ have masses
\be
|n_i\rangle :\quad  m_{wind,n_i}^2 \equiv n_i^2 r^2g_s^{(10)}(M_P^{(d+2)})^2\\, .
\label{wind2a}
\ee 
whereas the $SL(2)$ singlet winding states with two non-vanishing winding number $n_1=n_2=n$ have masses
\be
|n\rangle : \quad m_{wind,n}^2 \equiv 2 n^2 r^2g_s^{(10)}(M_P^{(d+2)})^2\, .
\label{wind2a}
\ee

%\vskip0.3cm

\subsection{Two dimensional orbifold}

In the following we are looking for an orbifold, for which the only the $SL(2)$ KK modes are invariant.
This  orbifold is defined as 
\begin{equation}
{\cal O}^{(2)}_A=T^{(2)}/\Gamma^{(2)}\, ,
\end{equation}
where $\Gamma^{(2)}_A$ is the ${\mathbb Z}_2$ group, which acts by interchanging the two coordinates  of the torus, i.e. 
\begin{equation}
\Gamma^{(2)}_A:\quad x_1\longleftrightarrow x_2\, .
\end{equation}
The group $\Gamma^{(2)}$   acts on the complex structure modulus $U$ as $U\rightarrow 1/\bar U$. Therefore, 
diving out by $\Gamma^{(2)}$, one has to set $r_1=r_2$, i.e. the  $U$ modulus is restricted to satisfy $U=e^{-i\theta}$. 
%This means that the torus has to be rectangular with equal radii. 
%Furthermore  only the $SL(2)$ singlet states $|l\rangle $ are invariant under this projection.
While the volume of the original torus is ${\cal V}(T^2)=r^2\cos\theta$, the orbifolded version ${\cal O}^{(2)}$ has a reduced volume: 
\begin{equation}{\cal V}({\cal O}^{(2)})={\cal V}(T^2)/2=(r^2\cos\theta)/2\, .
\end{equation}
Looking at the induced topology of the orbifold, the first homology group $H_1$ as well as the fundamental group $\pi_1$ of ${\cal O}^{(2)}$ are also reduced compared to the two-torus and are given as
\begin{equation}
H_1({\cal O}^{(2)}_A,{\mathbb Z})=\pi_1({\cal O}^{(2)})={\mathbb Z}\,.
\end{equation}
Moreover ${\cal O}^{(2)}_A$ has one 1-dimensional boundary, namely the line $x_1=x_2$, which is the fixed line of $\Gamma^{(2)}$.
In fact,  ${\cal O}^{(2)}_A$ is an non-orientable surface with $S^1$ as boundary, it is just two two-dimensional M\"obius strip.

The leading KK tower on the orbifold must be invariant under the $\Gamma^{(2)}_A$ group action; as said before it is the $SL(2)$ singlet tower $|l\rangle $ with $l_1=l_2=l$.
One can show  that the mass scale of this leading KK tower  is identical to inverse orbifold volume, only if one chooses $\theta=0$, i.e. if the underlying torus is rectangular.
Then the leading KK mass scale on the orbifold A can be indeed rewritten as
\be
m_{KK}^2         = {2\over r^2}=   {1\over {\cal V}({\cal O}^{(2)}) }(M_P^{(d+2)})^2\quad {\rm for}\quad\theta=0\, .
\label{KKT2dlbb}
\ee

\subsection{D-dimensional torus}

Next we generalize this discussion  to backgrounds ${\mathbb R}^{1,d-1} \times T^{(D)}$, with ${\mathbb R}^{1,d-1}$ being the $d$-dimensional Minkowski space and $T^{(D)}$ a torus of dimension $D$.
It is determined  by a constant metric $G_{ij}$, which is characterized by $D$ different radii $r_i$ ($i=1,\dots, D$) and $D(D-1)/2$ different angles $\theta_{ij}$.
The volume of $T^{(D)}$ is given as
\begin{equation}
{\cal V}(T^{(D)})=\sqrt{\det G_{ij}}\, .
\end{equation}
%which we first take  to be just the product of $D$ circles of same radius $r$. 
In general the mass spectrum  of the KK states has the following form
\begin{equation}
m_{KK}^2=l_iG^{ij}l_j(M_P^{(d+D)})^2\, ,
\end{equation}
where $G^{ij}$ is the inverse metric of the D-dimensional torus.
%In general the metric of the torus is characterized by $D$ different radii $r_i$ ($i=1,\dots, D$) and $D(D-1)/2$ different angles $\theta_{ij}$.
Moreover, certain discrete $SL(D)$ transformations on the $r_i$ and the $\theta_{ij}$, which build a subgroup of the general duality  group $SO(D,D,{\mathbb Z})$, leave  the torus unchanged. In 
order to keep the KK spectrum invariant, these transformations must be accompanied by corresponding linear transformation among the  KK quantum numbers $l_i$.

The leading KK towers $|l_i\rangle $ on the torus are the ones with a single non-vanishing KK charge $l_i$ ($i=1,\dots ,D$)
in  the i-th. direction. They transform in the fundamental representation $\underline D$ of the group $SL(D)$ and 
have a mass in terms $(d+D)$-dimensional Planck mass $M_{P}^{(d+D)}$
\be
%{
%\color{red}
|l_i\rangle :\quad m_{KK,i}^2 
%=\left(\frac{n_i}{R}\right)^2
=l_i^2 G^{ii}(M_P^{(d+D)})^2\, .
\label{KKi}
\ee

Next let us consider the subleading, $SL(D)$ singlet KK tower  $|l\rangle$ with all momenta equal to each other, i.e. $l_i=l$,
where the subleading tower mass scale in units of $M_{P}^{(d+D)}$  is given as
\be
|l\rangle :\quad m_{KK,l}^2 =   l^2  m_{KK,(S)}^2\, \quad{\rm with}\quad m_{KK,(S)}^2= \sum_{i,j}G^{ij}(M_P^{(d+D)})^2\,.\label{invariantA}
\ee
As we will see in the following, these masses depend on the number of dimensions $D$ in a particular, non-trivial way.

\subsection{D-dimensional orbifold}

Now we will consider  orbifolds of the form
\begin{equation}
{\cal O}^{(D)}=T^{(D)}/\Gamma^{(D)}\, .
\end{equation}
Here  $\Gamma^{(D)}$ 
is a certain discrete symmetry group of the D-dimensional torus.
The volume of the orbifold is in general given as
\begin{equation}
{\cal V}({\cal O}^{(D)})={\cal V}(T^{(D)})/|\Gamma_D|\, ,
\end{equation}
where $|\Gamma_D|$ is the order of the group $\Gamma^{(D)}$.
As we will see, only the towers that are subleading on the torus will be invariant under the orbifold group and will survive the orbifold projection.
So these towers will become the leading KK towers on the orbifolds and they indeed possess a non-trivial dependence on the number of dimensions.

%only the $SL(D)$ singlet states $|l\rangle $ are invariant.

\vskip0.3cm\noindent
Let us consider look for the   D-dimensional orbifold, which leaves the singlet KK tower invariant. For that
we take a torus which is symmetric under the exchange $i\leftrightarrow j$, which implies that all diagonal metric elements are equal to each other, i.e. $G_{ii}=r^2$ and also all
non-diagonal metric elements must be equal to each other, i.e. $G_{ij}=G_{ik}=G_{kj}=r^2\sin\theta$.
In this case 
 the volume of the torus becomes
\begin{equation}
{\cal V}(T^{(D)})=\sqrt{\det G_{ij}}={r^D}\sqrt{(1-\sin\theta)^{D-1}(1+(D-1)\sin\theta)}
\end{equation}

The orbifold group $\Gamma^{(D)}$ is the symmetric group $S_D$ of all possible permutations among the torus coordinates:
\begin{equation}
\Gamma^{(D)}:\qquad x_i\longleftrightarrow x_j\, ,\quad i,j=1,\dots ,D\, .
\end{equation}
Its order is given as $|\Gamma^{(D)}|=D!$ and the orbifold volume becomes
\begin{equation}
{\cal V}({\cal O}^{(D)})={\cal V}(T^{(D)})/D!=\frac{r^D}{D!}\sqrt{(1-\sin\theta)^{D-1}(1+(D-1)\sin\theta)}\, \label{orbivolA}
\end{equation}
We see that ${\cal V}({\cal O}^{(D)})$ decreases in the large D limit when keeping the diameter $r$ fixed, compared to the volume of the covering torus.

The orbifold action  leaves only one 1-cycle of the underlying torus invariant and hence
the first homology group $H_1$ and the fundamental group $\pi_1$ of ${\cal O}^{(D)}$ are given as
\begin{equation}
H_1({\cal O}^{(D)},{\mathbb Z})=\pi_1({\cal O}^{(D)})={\mathbb Z}\,.
\end{equation}
Furthermore the local structure of ${\cal O}^{(D)}$ is  of the form
\begin{equation}
{\cal O}^{(D)}=I_{D-1}\times S^1\,,
\end{equation}
where $I_{D-1}$ is a (D-1)-dimensional space with one isolated point.
Hence ${\cal O}^{(D)}$ possesses one  isolated $S^1$, which is  fixed by $\Gamma^{(D)}$ with $x_1=x_2=\dots =x_D$ and is a D-dimensional generalization of the
two-dimensional M\"obius strip.

Now the $SL(D)$ singlet KK tower  $|l\rangle$ of eq.(\ref{invariantA}) with all momenta equal to each other, i.e. $l_i=l$, becomes the leading tower on 
${\cal O}^{(D)}$;
it is
invariant under the orbifold group $\Gamma^{(D)}$ and survives
as invariant KK tower on the orbifold. So let us compare the corresponding KK masses in eq.(\ref{invariantA}) with the orbifold volume ${\cal V}({\cal O}^{(D)})$ in eq.(\ref{orbivolA}).
For the torus with $G_{ii}=r^2$ and  $G_{ij}=G_{ik}=G_{kj}=r^2\sin\theta$
this mass scale becomes
\be
m_{KK,(S)}^2 =  m_{KK,(S)}^2 =\frac{D}{r^2(1+(D-1)\sin\theta)}\,.\label{invariantABa}
\ee
We can now solve $m_{KK,(S)}^2={\cal V}({\cal O}^{(D)})^{-2/D}$ directly for any $D$, and we can check that indeed there exists a solution $\theta_D$ for any $D$, which for large $D$ behaves such that
\beq
\sin\theta_D\sim -\frac{1}{D}
\eeq
So for a special value of $\theta_D$ the KK mass scale can be indeed expressed in terms of the orbifold volume:
\be
m_{KK,(S)}^2 =   \biggl({1\over{\cal V}({\cal O}^{(D)})}
\biggr)^{2\over D}={1\over r^2} \biggl(
{D!\over \sqrt{(1-\sin\theta_D)^{D-1}(1+(D-1)\sin\theta_D)}}
\biggr)^{2\over D}\,.\label{invariantAB}
\ee
In  terms of the effective d-dimensional Planck mass $M_{P}^{(d)}={\cal V}({\cal O}^{(D)})^{1/(d-2)}M_{P}^{(d+D)}$,
the masses $m_{KK,(S)}^2$ can then be  expressed as
\be
 m_{KK,(S)}^2 
%=\left(\frac{n_i}{R}\right)^2
=   \biggl(
{D!\over r^D\sqrt{(1-\sin\theta_D)^{D-1}(1+(D-1)\sin\theta_D)}}
\biggr)^{2(D+d-2)\over D( d-2)} \,.
\label{KKiaaalarge}
\ee
One sees that for large $D$ and fixed $r$, these masses become very big. 
Note that this behaviour is very similar to the KK masses on the sphere - see \cite{Bonnefoy:2020uef}.

\vskip0.2cm
\noindent
Conversely 
$m_{KK,(S)}^2 \leq 1$ for D being smaller than a critical value, namely
\begin{equation}
D! \leq \sqrt{\det G_{ij}}=
r^D\sqrt{(1-\sin\theta_D)^{D-1}(1+(D-1)\sin\theta_D)}
\, .\label{Dboundtorus11}
\end{equation}
For large $D$ we can approximate $D! \approx D^D$ and $\theta_D=0$, i.e. $\det G_{ij}\approx r^{2D}$, where $r$ is the typical length scale of the torus, and then $m_{KK,(S)}^2 $ grows with $D$ as
\be
 m_{KK,(S)}^2 
%=\left(\frac{n_i}{R}\right)^2
\approx  \biggl({ D\over r}\biggr)^{2(D+d-2)\over d-2} \,.
\label{KKiaaalarge11}
\ee
Constraining $m_{KK,(S)}^2$ to be smaller than the Planck mass, i.e.  $m_{KK,(S)} \leq 1$, one derives the following simple bound on $D$:
\begin{equation}
D\leq r\, .\label{DboundtorusA}
\end{equation}
Note that this is a relation between $D$ and the radius of the original torus, which is indeed a physically meaningful relation, since the one-dimensional length scale of the orbifold is still given by $r$ and only its volume is reduced compared 
to the torus volume.

\section{Large Dimension Conjecture  and D-duality for orbifolds}\label{Dduality}

As we have seen in the previous section, the leading, invariant  KK states on the orbifold  ${\cal O}^{(D)}$
possess are non-trivial dependence on the number of dimensions $D$ and are approximately related to the following mass scale:
\be
 m_{KK,(S)}^2 \equiv m(D,r)=    \biggl({D^\alpha\over r^2} \biggr)               \, ,
 \ee 
with $\alpha $ being  in general a background dependent parameter, $\alpha=2$ for the orbifold A and $\alpha=1$ for the orbifold B.
%On the orbifold ${\cal O}^{(D)}$ this is indeed the unique KK tower.
 %The key question therefore is what could possibly happen in the quantum gravity
 %theory in case the LDC is violated? 
 As we discussed in the last section, 
 for 
 \begin{equation}
 D^\alpha>r^2\, ,\label{dbounda}
 \end{equation}
this mass scale of the leading  KK tower    becomes  heavier than the Planck scale and therefore all leading  KK states decouple from the theory at scales below $M_P$.\footnote{In 
case there are elementary winding states, like on the torus, the  T-dual tower of singlet elementary winding states $|n\rangle $ becomes  heavier than the Planck scale for $D^\alpha>1/R^2$.
 So irrespective of the perturbative T-duality symmetry between the  KK and winding states there is always a regime where both towers $|l\rangle $ and $|n\rangle$ become heavier than the Planck scale, namely
 \begin{equation}
 D^\alpha>\max\bigl(r^2,{1\over r^2}\bigr)\, .
 \end{equation}}

 Now let us assume that  the LDC (or the NDC) must hold for   the leading KK towers on  the orbifolds (or even for the non-leading towers on the torus).
 Then there are two options:
 either the relation (\ref{dbounda}) provides an upper bound on the allowed dimensions  or there must be a new kind of (non-perturbative) duality symmetry, called D-duality, that acts on $D$ and on $r$ in a non-trivial way.

 In the following  we want to investigate the necessary conditions that  D-duality in general gravity theories, a priori not restricted to string theory, can be at least in principle realized on the considered orbifolds.
 However for superstring theory resp.  for M-theory the number of dimensions is restricted to be smaller than 11, and for higher-dimensional constructions further truncations of the spectrum appear to be
 necessary.
  Concretely we want to investigate the possibility that  there exists a new tower of brane states $|\tilde l\rangle$  with masses $\tilde m$, possibly in a different number of dimensions $\tilde D$, 
 which is dual to the KK states in $D$ dimensions.
 Whereas the KK states become heavy for large $D$ and small $r$, the dual brane states become light for large $\tilde D$ and small $\tilde r$ and vice versa.
 %Moreover, if the dual states are completely equivalent to the original KK modes, then
 % inequivalent  theories are be already covered by the range  $(\sqrt D/r)<1$.
  Dual to the branes in $ D$ dimensions there should be also the corresponding  KK states in $\tilde D$ dimensions.
  The  wrapped brane  states in general correspond 
  to
  the  homotopy class $\pi_{ D}(  {\cal M})$ ($\pi_{\tilde D}( \tilde {\cal M})$) of the D($\tilde D)-$dimensio\-nal space. A further necessary condition 
for the existence of a full tower of one-particle wrapped brane states is that  the fundamental group of $ { M}$ ($ \tilde{\cal M}$) is non-trivial,
i.e. $\pi_{1}( {\cal M}^{( D)})\neq 1$ ($\pi_{1}( \tilde{\cal M}^{( \tilde D)})\neq 1$).
As we have seen, for both  D-dimensional orbifold spaces A and B under consideration we have that 
\begin{equation}
\pi_{1}( {\cal O}^{(  D)})={\mathbb Z}\, ,
\end{equation}
and hence one can  indeed get  a full tower of wrapped membrane states.

Since, as we have just seen,  the necessary topological condition for a dual tower of wrapped branes is satisfied,   a  D-duality symmetry is in principle possible, which then can act
 as:
  \begin{equation}
 {\rm D-duality}:\quad
| l\rangle \longleftrightarrow |\tilde l\rangle 
\end{equation}
with the following identification of KK and branes tower mass scales
\begin{eqnarray}
m_{brane}(D,r) &=& \tilde m_{KK} (\tilde D,\tilde r)\, ,\nonumber\\
 m_{KK}(D,r)& =& \tilde m_{brane} (\tilde D,\tilde r) 
 \,.\label{d-duality2} 
 \end{eqnarray}
% and possibly also vice versa.
 %Here $\alpha$ is a parameter, which will be determined in the following by the masses of the dual states $|\tilde l\rangle $ .
 
 Note that these relations are meant to hold in a given number of uncompactified dimensions $d$, which means that the effective theory in $d$ dimensions is invariant under D-duality. From a higher dimensional point of view,
 D-duality should be seen as duality 
 between two theories in $d+D$ and in $d+\tilde D$ number of dimensions.
For the same, fixed number of compact dimensions, namely $D=\tilde D$, D-duality is relating the radius $r$ with a dual radius $\tilde r$ in a particular way.
 Moreover we will also discuss the possibility that D-duality relates a theory in $D$ dimensions  with a theory in a dual, however different number of compact dimensions $\tilde D$; so it can possibly  act as 
   a  {\sl large D -- small D} duality transformation.

\section{Some concrete string examples of D-dual pairs}

Since string theory or M-theory is restricted by the existence of the critical dimension, the use of D-duality appears to be rather restricted. Nevertheless 
 we want to consider some dual pairs of theories in two different numbers of dimensions, namely string/M-theory duality and also string/F-theory duality.
 We will first start  in $d=9$ space-time dimensions. Following our general arguments, the KK states from $9+D$ dimensions must correspond to  wrapped branes
 in the dual
 $(9+\tilde D)$-dimensional geometry. 
 %For large $D$ the KK states in $9+D$ dimensions become heavy, whereas the corresponding branes in $9+\tilde D$ dimensions become light.
 The simplest example is the case with $D=2$ and $\tilde D=1$, which corresponds to  the well-known M-theory/string theory duality.
 For higher $ D$ one gets theories, which a priori possess more degrees of freedom than its dual string theory with $\tilde D$ compact dimensions. Therefore a truncation of the degrees of freedoms is necessary.
 An example for this is a 12-dimensional theory with $ D=3$ and a  3-brane being dual to type IIB in $d=9$ and $\tilde D=1$. As we will see, the necessary truncation is closely related to the section constrain  in exceptional field
 theory.

 \subsection{M-theory ($D=2$) and IIB string ($\tilde D=1$)}

  As it is well-known \cite{Witten:1995ex}, M-theory in 11 dimensions is  dual to IIA superstring in 10 space-time dimensions. Here we want to consider M-theory on a 2-torus and also on the 2-dimensional orbifold ${\cal O}^{(2)}$.
 The D-dual theory is IIB on $S^1$, which is obtained from type IIA on the circle plus one T-duality transformation. So following our previous notation, we have that $D=2$ and $\tilde D=1$ and
 we will now discuss the duality between IIB on $S^1$ and M-theory on $T^{(2)}$ from the D-duality perspective, possibly with a further ${\mathbb Z}_2$ projection on  ${\cal O}^{(2)}$.

 \subsubsection{Toroidal case}

 \vskip0.2cm
 \noindent

 The type IIB U-duality group in 9 dimensions is $SL(2,{\mathbb R})\times {\mathbb R}$ and the well-known 9-dimensional bosonic field content is as follows:
 
 \noindent
 (i) three scalar fields, which parametrize the moduli space $SL(2)/U(1)\times {\mathbb R}$, 
 
  \vskip0.2cm
  \noindent
 (ii)
 two vector fields $(A^{(1)}_{\mu  },  B^{(1)}_{\mu }   )  $, which transform as $\underline 2$ under $SL(2)$ plus one vector field $C^{(1)}_{\mu }$, which is a singlet under $SL(2)$,
 
  \noindent
(iii)  two 2-forms $(A^{(2)}_{\mu\nu },B^{(2)}_{\mu\nu })$,
  which transform as $\underline 2$ under $SL(2)$, 
  
   \noindent
  (iv)
   one 4-form $A^{(4)}_{\mu\nu\lambda \rho}$,
  which transform as $\underline 1$ under $SL(2)$. 
  
   \vskip0.2cm
   \noindent
All these fields can be obtained from the reduction of 11-dimensional M-theory metric $G_{mn}$ and the M-theory 3-form $C_{mnp}$ on $T^2$ plus one T-duality transformation in the well-known way \cite{Schwarz:1995dk}.

  \vskip0.2cm
   \noindent
 The corresponding  particles and branes in 9 dimensions are as follows (see also \cite{AbouZeid:1999fv}):
 
  \vskip0.2cm
   \noindent
  (i)  One D3-brane: it is the brane which is dual to the un-wrapped M2-brane of M-theory.
 
  \noindent
 (ii)  One D2-brane, which is the un-wrapped M2-brane.
 
  \noindent
 (iii)
 two strings, which transform as $\underline 2$ under the $SL(2)$ duality group. They are the  IIB D1- and F1-branes,  which the M2-branes wrapped around one direction of the $T^2$.
 Actually the tension of the general $(n_1,n_2)$-brane is given as \cite{Schwarz:1995dk}
 \begin{equation}
 T^{(n_1,n_2)}=\biggl(g_s^{(10)}n_1^2+{n_2^2\over  g_s^{(10)}}\biggr)^{1/2}M_s^2\,.
 \end{equation}

  \noindent
(iv)  one IIB KK particle: it corresponds to the M2-brane, completely wrapped around the compact$T^2$,  and transforming as singlet under the $SL(2)$ duality group.

 \noindent
  (v) two IIB particles, being wrapped F1 and D1  strings: they correspond to the two M-theory KK particles in the $x_9$ and $x_{10}$ directions,  and transforms as $\underline 2$ under the $SL(2)$ duality group.

 %\vskip0.2cm
% \noindent
 \vskip0.2cm
   \noindent
For illustrative reasons,  let us consider some of the  well known mass relations for the toroidal case. They nicely fit into the general D-duality framework.
 First we compare the masses of the $SL(2)$ singlet states, namely the IIB KK modes with the M-theory wrapped M2 modes. For simplicity we will set the two radii 
 of the M-theory 2-torus equal to each other, i.e. $ R_1= R_2= R$. Therefore the IIB string coupling $g_s^{(10)}=R_2/R_1$ is equal to one.
 The first D-duality relation in (\ref{d-duality2}), expressed in units of the 
 9-dimensional Einstein frame,  then becomes
  \begin{equation}
  m_{M2}^{(M)}= \tilde m_{KK}^{(IIB)}\quad\Longleftrightarrow\quad
 r^{12\over 7}={1\over  \tilde r ^{8/7}}
 \, .\label{d-duality6}
 \end{equation}
 
 Second we consider the $SL(2)$ doublet states, namely the two  KK states on the M-theory side and the two wrapped 1-branes on the IIB side. 
  Setting their masses equal to each other we get the second D-duality relation in (\ref{d-duality2}):
    \begin{eqnarray}
 F1:\qquad m_{KK}^{(M)}= \tilde m_{F1}^{(IIB)}\quad&\Longleftrightarrow&\quad
{1\over  r}\biggl({ 1\over   r }\biggr)^{2/7}=\biggl( {g_s^{(10)}}   \biggr)^{1/2}\tilde r^{6\over 7}
  \, ,\nonumber\\
  D1:\qquad m_{KK}^{(M)}= \tilde m_{D1}^{(IIB)}\quad&\Longleftrightarrow&\quad
{1\over  r}\biggl({ 1\over   r }\biggr)^{2/7}=\biggl( {g_s^{(10)}}   \biggr)^{-1/2}\tilde r^{6\over 7}
  \, .
   \label{d-duality9}
  \end{eqnarray}
From eq.(\ref{d-duality6})   one gets that $ r=\tilde r^{-2/3}$
Furthermore it  follows from eqs.(\ref{d-duality9})
 %that the 9-dimensional string coupling constant is determined  to be $g_s^{(9)}=r^{-4/7}$.
 %Via the relation 
 %$g_s^{(9)}=g_s^{(10)}/r^{4/7}$  one then obtains 
 that indeed $g_s^{(10)}=1$.
 One can check that  the two D-duality relations (\ref{d-duality6}) and (\ref{d-duality9})
 are of course identical to  the known relations between IIB and M-theory \cite{Witten:1995ex,Schwarz:1995dk}. So in this case D-duality is completely equivalent to the know duality between M-theory and the type IIB superstring
 in 9 dimensions.

  \subsubsection{Two-dimensional orbifold}

 \vskip0.2cm
 \noindent

Next
we want to consider the two orbifold version of  M-theory on $T^{(2)}$ and of IIB on $S^1$.\footnote{Lower dimensional heterotic M-theory orbifolds were considered in 
\cite{Dasgupta:1995zm,Faux:1999hm,Faux:2000dv}.}
In M-theory the $SL(2)$ corresponds to the exchange of the two directions of $T^{(2)}$,
so the $SL(2)$ is just the geometric modular group of $T^{(2)}$, which acts as $x_1\leftrightarrow x_2$ and also on the two radii as $ R_1\leftrightarrow  R_2$.
%Recall that the U-duality group of the 9-dimensional IIB is given by $G=SL(2)$. 
In IIB the group $SL(2)$ is the S-duality group
%respectively U-duality group 
of the compactified IIB theory, which acts on the type IIB axion-dilaton field $\tau$  as $\tau\leftrightarrow -1/\tau$.
Therefore on the type IIB side the ${\mathbb Z}_2$ orbifold
${\cal O}^{(2)}$ corresponds to a non-geometric operation, namely its corresponds to modding with the IIB S-duality 
%or respectively U-duality 
symmetry. 
Namely modding out by $SL(2)$, and
setting  $R= R_1= R_2$, means that the IIB string coupling is not a free parameter, i.e.  $g_s^{(10)}= R_1/ R_2=1$.
So the resulting orbifold theory is a certain S-fold 
%or U-fold 
of type II B.
%Modding out by $SL(2)$,
%namely setting  $\tilde R=\tilde R_1=\tilde R_2$, means that the IIB string coupling is not a free parameter, i.e.  $g_s^{(10)}= R_1/ R_2=1$.
%where   $c$ is a constant of order one as we will see below below.
%So the S-fold  is a theory at strong coupling. 

Only the $SL(2)$ singlet states survive the ${\mathbb Z}_2$ orbifold projection.
These are first  the IIB KK modes on $S^1$, 
%whereas the  IIB $SL(2)$ doublet  is given in terms of the F1 string and the D1 string,
% both wrapped on $S^1$.
whereas in M-theory, the $SL(2)$ singlet state is the wrapped M2-brane. 
%On the other hand, the two M-theory KK modes build a $SL(2)$ doublet and correspond to the F1 string and the D1 string in type IIB.
% Therefore
%the $SL(2)$ singlet M2-brane states survive the orbifold projection and on the IIB side, the KK particles are invariant under the $SL(2)$ symmetries. 
Because the volume of the M-theory orbifold is reduced by half compared to the 2-torus, the first
D-duality relation (\ref{d-duality6})
 gets an additional factors of $\sqrt{ D}=\sqrt2$ and takes the form
  \begin{equation}
m_{M2}^{(M)}=   \tilde m_{KK}^{(IIB)}\quad\Longleftrightarrow\quad
 \biggl( {r\over \sqrt2}\biggr)^{12\over 7}={1\over \tilde  r ^{8/7}}
 \, .\label{d-duality7}
 \end{equation}

We can also consider the wrapped fundamental F1-string and the wrapped D1-string on the IIB side. 
%These two states form a doublet under $SL(2)$.
The $SL(2)$ invariant string state is given by the linear combination $|brane\rangle_{inv}={1\over \sqrt2}(|F1\rangle +|D1\rangle)$.
On the M-theory side the linear combination of the two KK states, $|KK\rangle_{inv} ={1\over\sqrt2}(|l,0\rangle+|0,l\rangle=|l,l\rangle)$, is $SL(2)$ invariant. 
We have to remember that the mass of the invariant M-theory KK modes  $|KK\rangle_{inv}$  gets an additional factor of $\sqrt{ D}=\sqrt2$  and we also have to take into account the reduced orbifold volume when going to the 9-dimensional Einstein frame.
Then 
setting the brane  mass equal to KK mass we get the second D-duality relation
% \begin{equation}
%  m_{brane_{inv}}^{(IIB)}=\tilde m_{KK_{inv}}^{(M)}\quad\Longleftrightarrow\quad
%  \sqrt{g_s^{(10)}} r^{6\over 7}
%  ={\sqrt 2\over \tilde r}\biggl({ \sqrt2\over \tilde  r }\biggr)^{2/7}
%  \, .\label{d-duality8}
% \end{equation}
% From eq.(\ref{d-duality7})   we get that $\tilde r=\sqrt2r^{-2/3}$ and
 %from eq.(\ref{d-duality8})  one obtains that $g_s^{(10)}=1$.
  \begin{equation}
m_{KK_{inv}}^{(M)}=  \tilde m_{brane_{inv}}^{(IIB)}\quad\Longleftrightarrow\quad
{\sqrt 2\over  r}\biggl({ \sqrt2\over   r }\biggr)^{2/7}={1\over\sqrt2}\biggl( \sqrt{g_s^{(10)}}+{1\over  \sqrt{g_s^{(10)}}}    \biggr)^{1/2}\tilde r^{6\over 7}
  \, .\label{d-duality8}
 \end{equation}
 Note that due to the orbifold action the
 string tension in the  IIB orbifold is reduced by a factor of $1/\sqrt2$.
From eq.(\ref{d-duality7})   we get that $ r=\sqrt2\tilde r^{-2/3}$ and
 from eq.(\ref{d-duality8})  one obtains that again $g_s^{(10)}=1$.

Let us discuss what this orbifolded version of IIB or, respectively, of M-theory is. 
From the 11-dimensional perspective there is one 10-dimensional boundaries, where additional matter fields must be located. As said before the 
orbifold  ${\cal O}^{(2)}$ is just the non-orientable two-dimensional M\"obius strip. Actually M-theory compactified on the M\"obius strip was already discussed before in \cite{Aharony:2007du}.
There is a gauge theory with reduced rank=10 located at the boundary of the M\"obius strip. 
From the dual IIB perspective it is a strongly coupled  S-fold with the type IIB coupling  set equal  to one. To be more precise the dual theory corresponds
to a heterotic string theory, namely it is 
the socalled CHL string \cite{Chaudhuri:1995fk,Chaudhuri:1995bf}
in nine dimensions, which possesses one point in moduli
space with an unbroken $E_8$ gauge symmetry.
Then, 
the fixed  coupling constant of the type IIB S-fold corresponds in the heterotic description to a fixed angle $\theta$ of the M-theory torus.
So we see that the type IIB S-fold has a nice interpretation in terms of the CHL heterotic string.

\vskip0.3cm

%Second, the ${\mathbb Z}_2$ projection completely breaks space-time supersymmetry, since all 32 supercharges are not invariant under this projection ({\bf check}).
%Third, the ${\mathbb Z}_2$ transformation $x_{10}\leftrightarrow x_{11}$ possesses one 1-dimensional boundary in the $T^2$, namely the line $x_{10}= x_{11}$.
%This is still  very reminiscent of the M-theory heterotic string construction mentioned before.
%Namely it is conceivable that the non-supersymmetric ${\mathbb Z}_2$-orbifold with one boundary, considered here, corresponds to the non-supersymmetric heterotic string with only one $E_8$ gauge group.
%The spectrum of this theory only contains space-time bosons and no fermions.
%The heterotic dilaton is fixed by its potential to its string coupling value.

One can also consider analogous D-dual pairs in a smaller number $d$ of uncompactified dimensions by increasing the number of compact dimensions,
e.g.  
with $d=4$, $D=7$ and $\tilde D=6$ or with $d=2$, $D=9$ and $\tilde D=8$.
In this case the orbifold action will lead to some exotic S-folds or U-folds,
similar to the type II orbifolds, which were constructed in  \cite{Florakis:2017zep,Ferrara:2018iko}.
It would be interesting to see, if these S- or U-folds again have a simpler interpretation as a compactified CHL-like heterotic string theory.
 Since the associated orbifolds are  non-orientable and compactifications of this kind lead to orentifolds without vector structure 
 \cite{Witten:1997bs,Bachas:2008jv}.

 % \subsubsection{$\tilde D=D+1$}

  \subsection{12-dimensional theory ($ D=3$) and IIB string ($\tilde D=1$)}

   \subsubsection{Toroidal case}

 Now we want to go towards larger $D$ by increasing the number  of dimensions $D$ by one and keeping  $\tilde D=1$.
 For that we consider again type IIB on ${\mathbb R}^{1,8}\times S^1$. The dual theory with $ D=3$ is  a 12-dimensional theory,
 being closely related to F-theory, compactified on $T^3$  or an orbifold version of it. 
 D-duality now requires the existence of a fundamental 3-brane in 12 dimensions. It
 will turn out to be the D3-brane of the IIB superstring. 
 Due to the close relation to F-theory \cite{Vafa:1996xn}, we will call it the F3-brane and it can be wrapped around the entire 3-dimensional compact space and then leads to a full tower of particles in 9 dimensions. 
  This tower of states transform as a singlet under the $SL(3)$ duality group.
 The corresponding, dual KK particles are the KK modes of type IIB on $S^1$.

 %A priori, the duality group  of the 12-dimensional theory in 9 dimensions is given by the group $SL(3)$. In order to identify the compactified 12-dimensional theory with type IIB (or type IIA)
 %in 9 dimensions, the group $SL(3)$ must be broken to $SL(2)$.
 Let us investigate in more detail what kind of 12-dimensional theory on $T^3$ can be D-dual to 9-dimensional type IIB on ${\mathbb R}^{1,8}\times S^1$.
 A priori, the duality group  of the 12-dimensional theory in 9 dimensions is given by the group $SL(3)$.
 Therefore under dimensional reduction on $T^3$ all 12-dimensional fields are transforming under irreducible representations of $SL(3)$.
  In order to identify the compactified 12-dimensional theory with type IIB 
 in 9 dimensions, the group $SL(3)$ must be broken to $SL(2)$.
Hence we also decompose the  irreducible representations of $SL(3)$ 
 under its $SL(2)$ subgroup. 
 
 First we will assume that the 12-dimensional theory possesses a metric $G_{MN}$. It decomposes in 9 dimensions as, with  internal index $i=1,2,3$: 
 \begin{itemize}
 \item
 9-dimensional metric $G_{\mu\nu}$,
 \item
 3 vector fields $G^{(1)}_{\mu i}$ ($i=1,2,3$), which transform as $\underline 3$ of $SL(3)$ and as $\underline 1+\underline 2$ under $SL(2)$, 
 \item
  6 scalar fields $G^{ij}$, which transform as $\underline 2+\underline 4$  under $SL(2)$, and which parametrize the  geometric moduli space ${\cal M}=(SL(3)\times {\mathbb R})/SU(2)$ of $T^3$.
 \end{itemize}
 Second, 
 the existence of the wrapped F3-branes implies   an Abelian $4$-form $D^{(4)}_{MNPQ}$ with corresponding $5$-form field strength $H^{(5)}_{MNPQR}$. In 9 dimensions it decomposes as follows:
 \begin{itemize}
  \item
  one singlet vector field  $D^{(1)}_{\mu ijk}$,
   \item
  three 2-forms $D^{(2)}_{\mu\nu ij}$,
  which transform as  $\underline {\bar 3} $ of $SL(3)$ and as $\underline 1+\underline 2$ under $SL(2)$,
   \item
 three 3-forms $D^{(3)}_{\mu\nu\lambda i}$,
  which transform as  $\underline 3$ of $SL(3)$ and as $\underline 1+\underline 2$ under $SL(2)$,  
 \item
 one 4-form $D^{(4)}_{\mu\nu\lambda\rho}$, which is a singlet under $SL(2)$ (and which is dual to a 3-form in 9 dimensions).
 \end{itemize}
  Together with the metric these comprise 264 bosonic degrees of freedom.

 Let us connect these 12-dimensional fields to the bosonic fields of  type IIB  in 9 dimensions. Specifically the type IIB fields can be expressed in terms of the F-theory fields in the following way, where we spit the internal space
 index now as $i=(\alpha,s)$ with $\alpha=1,2$ and $s=3$:
\begin{itemize}
 \item 
 three scalar fields $\phi=G_{2,2}$, $g_{1,1}=G_{1,1}$ and $A=G_{1,2}$, which transform as $\underline 1+\underline 2$ under $SL(2)$, and which parametrize the moduli space $SL(2)/U(1)\times {\mathbb R}$,
 \item
 two vector fields $(A^{(1)}_{\mu  },  B^{(1)}_{\mu }   )    =G_{\mu \alpha}$, which transform as $\underline 2$ under $SL(2)$ plus one vector field $C^{(1)}_{\mu }=D^{(1)}_{\mu \alpha\beta s}$, which is a singlet under $SL(2)$,
 \item
 two 2-forms $(A^{(2)}_{\mu\nu },B^{(2)}_{\mu\nu })=D^{(2)}_{\mu\nu \alpha s}$,
  which transform as $\underline 2$ under $SL(2)$, 
  \item
   one 4-form $A^{(4)}_{\mu\nu\lambda \rho}=D^{(4)}_{\mu\nu\lambda\rho}$, which is dual to a 3-form and
  which transform as $\underline 1$ under $SL(2)$. 
 \end{itemize}
 These  fields indeed comprise the  128 bosonic fields of type IIB. %Comparing with the total number of fields in 12 dimensions, we see that the type IIB string contains 136 fields less. 
 We see that the 12-dimensional theory contains 
 136 additional bosonic  degrees of freedom compared to the type IIB string. 
 %Concretely the 136 additional fields are 3 scalars, one vector field $G_{\mu s}$, one 2-form field $D^{(2)}_{\mu\nu \alpha \beta}$
 %and three 3-form fields $D^{(3)}_{\mu\nu\lambda i}$. 
 In order to match 
  the 12-dimensional field
 content with the type IIB field content, one can first restrict the three-torus $T^3$ to be a direct product $T^2\times S^1$. This will reduce the number of massless scalars from 6 to 4.
 In addition, like in F-theory, the volume modulus of the $T^2$ must be frozen, such that only the complex structure modulus of the 2-torus will survive.
 Furthermore one has to set to zero 
  the vector field $G_{\mu s}$, the 2-form field $D^{(2)}_{\mu\nu \alpha \beta}$
 and the three 3-form fields $D^{(3)}_{\mu\nu\lambda i}$. 
 
 After this truncation the theory does not allow anymore for a full 12-dimensional geometric interpretation. Actually
 one ends up with the field content that was already discussed in the context of exceptional $SL(2)$ field theory in 12 dimensions by  \cite{Berman:2015rcc}. 
 As in  \cite{Berman:2015rcc} one can 
 impose a section condition
 of the following form
\begin{equation}
\partial_\alpha\partial_s=0\,.
\end{equation}
This condition has the effect that the remaining fields cannot depend on all  coordinates of the $T^3$. 
It possesses two solutions, namely one, which leads to M-theory or equivalently to type IIA, and a second one which leads to type IIB in nine-dimensions.
In our discussion, we have implicitly assumed the type IIB solution $\partial_\alpha=0$, namely that all fields do not depend on the two coordinates $x_i$. 

However note that our field content is 
different from the one introduced in  \cite{Berman:2015rcc}. Concretely, in \cite{Berman:2015rcc} three additional vector fields  in 9 dimensions were introduced, which transform as $\underline 1+\underline 2$ under $SL(2)$.
In our case these three vector fields originates from the 12-dimensional metric and the 12-dimensional 4-form - see above.

%As we will see below,  we can consider instead of the truncation an orbifolded version of the 12-dimensional theory, which possibly has a full 12-dimensional geometric interpretation.

%Perhaps write down action.

 \vskip0.5cm
 Let us now discuss the branes and the particles of the 9-dimensional type IIB theory. They can be all derived
 from the 12-dimensional F3-brane and from the momentum (KK) modes in the internal directions  in the following way:
 \begin{itemize}
 \item
One IIB D3-brane: it corresponds to the unwrapped F3-brane and transforms as singlet under the $SL(3)$ duality group.
 \item one IIB D2-brane: it corresponds to the F3-brane, wrapped around the compact $x_{12}$ direction,  and transforms as singlet under the $SL(2)$ duality group.
 \item IIB  D1- and F1 strings: they correspond to the two double-wrapped F3-branes, wrapped around the compact $(x_9, x_{12})$ and $(x_{10}, x_{12})$ directions,  and transform as $\underline 2$ under the $SL(2)$ duality group.
 \item one IIB KK particle: it corresponds to the F3-brane, completely wrapped around the compact$T^3$,  and transforms as singlet under the $SL(2)$ duality group.
  \item two IIB particles, being wrapped F1 and D1  strings: they corresponds to the two KK particles in the $x_9$ and $x_{10}$ directions,  and transforms as $\underline 2$ under the $SL(2)$ duality group.
 \end{itemize}

 %We can also combine the two R-R sectors of type IIA and type IIB plus the common NS-NS sector, which leads to the following field content

 It is clear that the full 12-dimensional theory on $T^3$ before the truncation possesses more branes and particles than its type IIB counterpart. In particular
 the full 12-dimensional theory leads to a $SL(3)$ triplet of KK particles along all directions of $T^3$, as well as to a $SL(3)$  triplet of double-wrapped F3-branes, which correspond to three 9-dimensional strings, denoted
 by F1, D1$_1$ and D1$_2$.

Now consider the D-duality mass relations between the 12-dimensional theory on $T^3$ and type IIB on $S^1$. First, the F3-brane, wrapped around the entire $T^3$, leads to a tower of particles in 9 dimensions. 
  This tower of states transform as a singlet under the $SL(3)$ duality group.
 The corresponding, dual KK particles are the KK modes of type IIB on $S^1$.
  The mass scale of the wrapped F3-branes in 9-dimensional Planck units is given as:
  \begin{equation}
  m_{F3}^{(F)} =
 r^{18\over 7}
 \, .\label{f-duality1}
 \end{equation}
 Hence the first duality relation in (\ref{d-duality2}) becomes
  \begin{equation}
  m_{F3}^{(F)}= \tilde m_{KK}^{(IIB)}\quad\Longleftrightarrow\quad
  r^{18\over 7}={1\over  \tilde r ^{8/7}}
 \, .\label{f-duality2}
 \end{equation}

 Second we consider the KK states on the F-theory side with the wrapped 1-branes on the IIB side. 
  Setting their masses equal to each other we get the second D-duality relation in (\ref{d-duality2})
 \begin{equation}
m_{KK}^{(F)}=  \tilde m_{brane}^{(IIB)}\quad\Longleftrightarrow\quad
 \biggl({1\over   r }\biggr)^{10/7}=\sqrt{g_s^{(9)}} \tilde r^{8\over 7}
  \, .\label{f-duality3}
 \end{equation}
% Using $g_s^{(9)}=1$
 %, the last relation becomes
  %\begin{equation}
 %r^{8\over 7}
 % =\sqrt 2\biggl({\sqrt 2\over \tilde  r }\biggr)^{9/7}
 % \, .\label{d-duality9}
% \end{equation}
From eq.(\ref{f-duality2})   one has that $ r=\tilde r^{-4/9}$.
Furthermore it  follows from eqs.(\ref{f-duality3}) that the 9-dimensional string coupling constant is determined  to be $g_s^{(9)}=r^{-64/63}$.
 Via the relation 
 $g_s^{(9)}=g_s^{(10)}/\tilde r^{4/7}$  one then obtains that  $g_s^{(10)}=r^{-4/9}$.

 \subsubsection{Orbifold case}

Now we want to describe the orbifolded version of the 12-dimensional ($D$=3)/9-dimensional ($\tilde D$=1)  duality.
Namely we consider the following orbifold:
\begin{equation}
{\cal O}^{(3)}=T^{(3)}/\Gamma_3\, ,
\end{equation}
where $\Gamma_3=S^3$ is the permutation group of six elements, which acts on the three coordinates as:
\begin{equation}
\Gamma^{(3)}:\qquad x_i\longleftrightarrow x_j\, ,\quad i,j=1,\dots ,3\, .
\end{equation}
The volume of ${\cal O}^{(3)}$ is given as 
\begin{equation}{\cal V}({\cal O}^{(3)})=\biggl({ r\over 6^{1/3}}\biggr)^3\, ,
\end{equation}
and the local structure of ${\cal O}^{(3)}$ is
\begin{equation}
{\cal O}^{(3)}=I_2\times S^1\,,
\end{equation}
where $I_2$ is a two-dimensional space with one isolated fixed point 
and with $H_1({\cal O}^{(3)},{\mathbb Z})=H_2({\cal O}^{(3)},{\mathbb Z})=\pi_1{\cal O}^{(3)}={\mathbb Z}$.
Hence ${\cal O}^{(3)}$ possesses one 1-dimensional one fixed cirle of $\Gamma^{(3)}$.

\vskip0.3cm
Under the orbifold projection only the $SL(3)$ singlet  towers are invariant.
These are:
 \begin{itemize}
 \item
The unwrapped F3-brane, which corresponds to a 3-brane in 9 dimensions.
 \item A $SL(3)$  invariant  single wrapped F3-brane around the fixed $S^1$ circle,  which leads to a tower of 2-branes in 9 dimensions. 
  \item A $SL(3)$  invariant, doubled wrapped F3-brane around $I_2$,  which leads to one 1-brane in 9 dimensions. 
 \item The completely wrapped F3 around the compact ${\cal O}^3$,  which corresponds to the tower of KK particles in 9 dimensions.
  \item One tower of  a $SL(3)$  invariant linear combination of KK particles. It corresponds to the $SL(3)$  invariant linear combination 
  of wrapped F1, D1  and D1' strings.
 \end{itemize}

The mass relations for the two sets of particles in 9 dimensions are
  \begin{equation}
  m_{F3}^{(F)}= \tilde m_{KK}^{(IIB)}\quad\Longleftrightarrow\quad
  \biggl({r\over 6^{1/3}}\biggr)^{18\over 7}={1\over  \tilde r ^{8/7}}
 \, ,\label{f-duality10}
 \end{equation}
and
 \begin{equation}
m_{KK}^{(F)}=  \tilde m_{brane}^{(IIB)}\quad\Longleftrightarrow\quad
\biggl({6^{1/3}\over r}\biggr)^{10/7}={1\over3^{1/3}}\biggl( 1+\sqrt{g_s^{(10)}}+{1\over  \sqrt{g_s^{(10)}}}    \biggr)^{1/3}\tilde r^{8\over 7}
  \, .\label{f-dualit11}
 \end{equation}

\vskip0.3cm
But since we started from 12-dimensional F-theory, 
a further truncation of the degrees of freedom should be applied.
The resulting theory in 9 space-time dimensions is an exotic,  S-fold like theory, whose precise relation to string theory is not obvious.
The orbifold action possesses one co-dimension-two fixed line, i.e. one  (9+1)-dimensional hypersurface.
Like for the heterotic M-theory, one expects that certain matter fields are located at on this hypersurface in 9 dimensions.
A priori the theory  contains additional 3-dimensional and 2-dimensional branes. In the simplest case, they can be eliminated by the standard F-theory truncation, described above. In this case one is ending again
at the same theory as M-theory on the M\"obius strip which is dual to the heterotic  CHL string in nine dimensions. If other consistent for the case of the three-dimensional orbifold ${\cal O}^{(3)}$ exist is 
 however not clear.

 \subsection{Higher dimensions}

Let us sketch how one can possibly proceed to higher dimensions, considering
 a dual  pair of compactifications with a compact space ${\cal M}$ of dimension $D$, being dual to $\tilde {\cal M}$ of dimension $\tilde D$.
$ {\cal M}$ and $\tilde {\cal M}$ are both orbifolds of the form
\begin{equation}
{\cal O}^{(D)}=T^{(D)}/\Gamma_D\, ,\quad{\rm and}\quad {\cal O}^{(\tilde D)}=T^{(\tilde D)}/\tilde\Gamma_{\tilde D}\, ,
\end{equation}
with the same orbifold types on both sides of the D-duality.
This is necessary for that the 
$SL(D)$ and $SL(\tilde D)$
 invariant states match up for the two orbifolds.

D-duality relates the KK modes in $d+D$ dimensions with 
the   branes that are wrapped around the entire compact space $\tilde {\cal M}$ of dimension $\tilde D$, and vice versa.
%These are $\tilde D$-dimensional branes with brane tensions $T^{(\tilde D)}$.  
 Concretely we assume that in $d+ D$ dimensional theory there exist an Abelian $(1+D)$-form $A_{(1+ D)}$ with corresponding $(2+ D)$-form field strength $F_{(2+ D)}$.
 This field strength is then sourced by a $ D$-dimensional brane, which can be naturally wrapped our the compact 
 ${\cal M}$. From the  point
 of view of the lower dimensional field theory in ${\mathbb R}^{1,d-1}$, the wrapped branes are particles, which couple to a 2-form gauge field strength $F_2$.
We assume  that the brane tension $T_{brane}^{( D)}$ is  determined by the higher-dimensional Planck
mass $ M_{P}^{(d+ D)}$ as
\be
T_{brane}^{( D)}=( g^{(d+ D)})^\alpha\bigl( M_{P}^{(d+ D)}\bigr)^{ D+1}\, .
\label{tension}
\ee
Here $( g^{(d+ D)})^\alpha$ is an elementary coupling constant in the $(d+ D)$-dimensional theory, related to a dilaton field, and $\alpha$ is an a-priori undetermined parameter.
The same kind of relations also hold in $d+\tilde D$ dimensions.

%{\bf Introduce coupling with certain power}

Now we are ready to employ the relations eq.(\ref{d-duality2})  between the KK modes of ${\cal O}^{(D)}$ and the winding modes of the orbifold ${\cal O}^{(\tilde D)}$ as necessary conditions for D-duality.
For the orbifolds at large dimensions
the duality relations take the form
  \begin{eqnarray}
  m_{brane}(D,r) &=& \tilde m_{KK} (\tilde D,\tilde r)\quad\Longleftrightarrow\quad
 ( g^{(d+ D)})^\alpha\biggl({ r\over  D}\biggr)^{D(d-3)\over d-2}=
    \biggl({{\tilde D}\over \tilde r} \biggr)^{\tilde D+d-2\over d-2}\, ,\nonumber\\
  m_{KK}(D,r)&=&\tilde m_{brane}(\tilde D,\tilde r)\quad\Longleftrightarrow\quad
   \biggl({ D\over r }\biggr)^{D+d-2\over d-2}
 =
 (\tilde g^{(d+\tilde D)})^\alpha\biggl({\tilde r\over{\tilde  D}}\biggr)^{\tilde D(d-3)\over d-2}
 \, .
  \label{d-duality21}
 \end{eqnarray}
  We see that   for a given pair of dual dimensions  
$ D$ and $\tilde  D$ the above equations imply particular relations between $r$ and $\tilde r$.
Note that the D-dimensional branes respectively the 
$\tilde D$-dimensional branes can be also wrapped around the invariant 1-cycle of the orbifolds, leading to further branes in the effective theory, with additional conditions on the mass spectra in the dual
theories. Furthermore, identification with critical superstring theories will require a truncation of the spectrum, similar to the F-theory cases, considered before.

Concerning string theory, the 12-dimensional case with $ D=3$ and $\tilde D=1$ can be  generalized to higher-dimensional  F-theory like constructions. E.g. a duality between type II in 8 space-time dimensions on $T^2$ $(\tilde D=2)$ and a 
13-dimensional S-theory with $D=5$ was discussed in \cite{Kumar:1996zx,Liu:1997mb,Curio:1998bv}. Following the D-duality proposal, 
it implies the existence of a 5-dimensional S-brane in 13 dimensions, wrapped around $T^5$,  which is dual to the singlet KK states on $T^2$.
In order to match this theory with string theory, the $SL(5)$ duality group must be broken to the 8-dimensional U-duality group of the type II string, which is $SL(3)\times SL(2)$.
Therefore a section condition must be chosen, which breaks  $SL(5)$ to $SL(3)\times SL(2)$. This can be partly achieved by restricting the five-torus $T^5$ to be a direct product $T^3\times T^2$.
The mass relations can be easily worked out using the equations given before.

\section{D-duality as strong-weak coupling duality}

As we will discuss now,  D-duality can be also regarded as  strong-weak coupling duality, now for a dual pair of theories on an orbifold with  a fixed, given number of D compact dimensions.
Consider first string theory on a circle circle compactification with Abelian gauge symmetry is $U(1)_L\times U(1)_R$. The KK modes are electrically charged under the diagonal subgroup $U(1)_{L+R}$
with electric gauge coupling constant $g^{(e)} \simeq 1/r$ , where the KK quantum number $l$ corresponds to their electric charges.
The winding modes are electrically charged under the orthogonal linear combination $U(1)_{L-R}$. Their  electric charges are given in terms of the winding numbers $n$, and
the electric gauge coupling constant  $\tilde g^{(e)} $ of the winding states is proportional to the radius,  $\tilde g^{(e)}\simeq r$.

Generalizing to toroidal  compactifications,
the effective theory on ${\mathbb R}^{1,d-1}$ possesses an Abelian gauge symmetry with gauge group $G=U(1)_L^{D}\times U(1)_R^{D}$
with electric 2-form field strengths $F_{2,i}^{(e)}$ coming from the internal part of the metric.
 The KK modes on ${\mathbb R}^{1,d-1}$ are charged with respect to the gauge symmetry group $G$ and
 the corresponding electric gauge couplings $g_i^{(e)}$ of the  $U(1)$ gauge symmetries are
 %in agreement with the weak gravity relation 
 proportional to the corresponding KK mass scales.

 Finally consider the orbifold effective field theory, and take for concreteness the A orbifold ${\cal O}^{(D)}=T^{(D)}/\Gamma_A^{(D)}$.
 Only one overall $U(1)$ gauge group with associated field strength $F_{2}^{(e)}$ remains as invariant gauge symmetry.
 The associated charged states are given by the leading KK tower 
  tower of the form $|1,1,\dots,1\rangle$.
    The corresponding gauge coupling is determined by  the KK mass scale in  eq.(\ref{KKiaaalarge}), which for large D becomes  
 \begin{equation}
 g^{(e)}\simeq \biggl({{ D}\over r}\biggr)^{D+d-2\over d-2}\, .\label{electricgauge}
 \end{equation}
 We see that the gauge theory becomes strongly coupled for $D\geq r$.
 Therefore asking for a weakly coupled gauge theory we derive the same constraint as from demanding that the associated KK tower is starting below the Planck mass.

The
  Abelian $U(1)$ gauge theory  is expected to
 possess a strong-weak coupling Olive/Mon\-to\-nen-like duality symmetry \cite{Montonen:1977sn,Font:1990gx,Schwarz:1993vs,Sen:1994fa}.
  %\footnote{For string theory on $AdS_5\times S^5$ this duality is not the same as the type IIB S-duality \cite{Schwarz:1995dk} that acts on the string coupling constant.}
 The corresponding dual magnetically charged states couple to the dual field strength $\tilde F_{d-2}^{(m)}$, which is a $(d-2)$ form on ${\mathbb R}^{1,d-1}$. 
 %The associated dual magnetic dual gauge coupling is given as
% \begin{equation}
 %g^{(m)}\simeq \biggl({r\over \sqrt D}\biggr)^\alpha\, ,
% \end{equation}
 %with some parameter $\alpha$, which we will determine in the following.
% 
 %
 %
%
 Actually, the magnetically charged objects originate from  a $({d}-4)$-brane in the effective ${\mathbb R}^{1,d-1}$ theory, which is the magnetic dual to the KK particles and is the magnetic source of  the $\tilde F_{d-2}^{(m)}$.   
 Lifting this form to the (d+D)-dimensional space, it corresponds to a magnetic $\tilde F_{d+D-2}^{(m)}$ form with a gauge potential $\tilde A_{d+D-3}^{(m)}$.
 It couples to a $({d}+D-4)$-brane, which is wrapped around the orbifold space.
  From the higher dimensional point of view, the magnetic $(d+D-4)$-branes are just  Kaluza-Klein monopoles, being  wrapped  around around
  the compact space. 
 The magnetic coupling $g_m$ is proportional to its  tension $T_m$, which in units of the d-dimensional Planck mass is given as
 %\footnote{Alternatively, the brane tension can be also derived from
% the weak gravity conjecture in arbitrary dimensions for arbitrary $p$-form gauge fields (with  coupling $g$) by reversing the extremality bound of charged black branes \cite{ArkaniHamed:2006dz,Heidenreich:2015nta}. The claim is that there should be a state of tension $T^{(m)}$ and quantized charge $Q$ such that
%\be
%\frac{p(D-p-2)}{D-2}T^2\leq g^2Q^2(M_P^{(D)})^{D-2} \ .
%\ee}
 \begin{equation}\label{tensionKK}
g^{(m)}\simeq T^{(m)}\simeq \biggl({ r\over  D}\biggr)^{D(d-3)\over d-2}\, .
 \end{equation}
 So from this perspective,  D-duality on the orbifold ${\cal O}^{(D)}$ is nothing else than the electric -- magnetic duality in the effective gauge theory on ${\mathbb R}^{1,d-1}$:
  \begin{equation}
 {\rm D-duality}:\quad g^{(e)}\,\, \longleftrightarrow \,\, g^{(m)}  \, .\label{d-dualityem}
 \end{equation}

 However the KK modes and the branes are  not self-dual objects, since they have in general different dimension. Therefore, the conjectured D-duality, namely the  electric-magnetic duality in ${\mathbb R}^{1,d-1}$,
  is in general not a self-duality symmetry but a duality between the electric, point-like KK modes
 and magnetic $(d-4)$-branes.
 
% A well-known string example of this scenario is the S-duality symmetry \cite{Font:1990gx,Schwarz:1993vs,Sen:1994fa} arising in the 10-dimensional heterotic string on the background ${\mathbb R}^{1,3}\times T^6$.
 %The electric gauge group in four dimensions is $U(1)^6$. The KK modes and their electric gauge couplings  are dual to the four-dimensional magnetic monopoles, which originate from  NS 5-branes being wrapped around
 %the 5-cycles of the six-torus. 

\section{Summary}

As we have discussed in this paper, the volume of certain D-dimensional compact orbifolds possesses a non-trivial dependence on D, namely it shrinks at large D.
This behaviour  
 is then inherited by the mass scale of the leading KK
tower in the associated effective field theory. This observation can be utilized to apply the Large Distance Conjecture (LDC) \cite{Bonnefoy:2020uef} in order to derive a general bound on the number of dimensions of orbifold
compactifications, which
is typically of the form
\begin{equation}
 {\rm LDC}-{\rm bound:}\qquad D^\alpha<r^2\, ,
 \end{equation}
with $\alpha$ being a background dependent parameter. This bound on $D$
 depends on the length scale $r$ of the theory.
 This suggests that 
in the IR for large scales there is a good notion of geometry, and the concept of a large number of space-time dimensions is well defined.
However in the UV at small distances for small $r$, the notion of geometry breaks down, and the number of space-time dimensions is bound by a number of order one.

As alternative of excluding the large $D$ regime of quantum gravity by the LDC-bound, one can investigate the possibility of a D-duality symmetry, which acts non-trivially on $D$ and $r$.
As we have discussed, the non-trivial fundamental homotopy group of the considered orbifolds allow for full towers of particles in the effective theory,  which is dual to the KK tower of states and which
is due to branes with (D+1)-dimensional world volume, being completely wrapped around the D-dimensional orbifold. M-theory on the M\"obius strip 
with a tower of wrapped M2-branes and its realization as CHL heterotic string in 9 space-time dimension
 is  a
non-trivial example of a D-duality based on an orbifold compactification. From the IIB perspective, this orbifold theory is a non-perturbative IIB S-fold, namely it corresponds to  an orientifold without vector
structure.

Going one step higher in the number of dimensions, we have investigated a D-duality between a 12-dimensional theory, closely related to F-theory and compactified
on a 3-dimensional space, and its D-dual theory on a circle from 10 to 9 dimensions. 
The 12-dimensional theory contains fundamental F3-branes, wrapped around the compact space, being dual to the KK modes of the dual circle.
The orbifold version of this dual pair leads to another rather exotic, non-supersymmetric 
 S-fold in 9 dimensions, with additional matter fields on the  orbifold fixed plane. It is still not clear, if this orbifold is a consistent quantum gravity theory or if it already belongs to the swampland.

Even much less is known about the existence of possible D-dual pairs originating from gravitational theories in dimensions higher than 12. 
Space-time supersymmetry is apparently not anymore possible in high number of dimensions, and therefore there are severe restrictions from stability considerations on the quantum consistency
of gravitational theories in the large D limit. This is similar to the bosonic string theory, and previous attempts 
\cite{Freund:1984nd,Casher:1985ra,Englert:1986na,Lust:1987ik}
to derive the superstring via compactification from the bosonic string can be also 
viewed from the point of view of D-duality, possibly extended to orbifold theories.
Finally it would interesting to see, if the swampland cobordism conjecture 
\cite{McNamara:2019rup,Montero:2020icj}
can provide further input for gravitational theories at large D.

\vskip1.5cm
\vspace{10px}
{\bf Acknowledgements}
\vskip0.1cm
\noindent
I like to  thank  Luca Ciambelli, Mariana Grana, Eran Palti, Cumrun Vafa and Timo Weigand
for very useful discussions
and in particular Quentin Bonnefoy and Severin L\"ust
for a lot of substantial input to the paper.
The work  is supported  by the Origins Excellence Cluster.

%For light soft modes the effective 2D theory is in the swampland and needs the completion to the 4D theory.
%%%%%%%%%%%%%%%%%%%%%%%%%%%%%%%%%%%%%%%%%%%%%%%%%%%%

%%%%%%%%%%%%%%%%%%%%%%%%%%%%%%%%%%%%%%%%%%%%%%%%%%%%

\end{document}